\documentclass[usenatbib,usegraphicx,usedcolumn]{mn2e}
\usepackage{times,amssymb,amsmath}
\usepackage{relsize}

\usepackage[backref,breaklinks,colorlinks,citecolor=blue]{hyperref}
\usepackage[usenames,dvipsnames]{color}
\def \mdot{\dot{m}}

\title[multi-accretion regime MAD]{Interpreting MAD within multiple accretion regimes}
\author[Mocz, Guo]{Philip Mocz$^{1}$\thanks{E-mail: pmocz@cfa.harvard.edu (PM)} and Xinyi Guo$^{1}$  \\
Harvard-Smithsonian Center for Astrophysics, 60 Garden Street, Cambridge, MA 02138, USA \\
}
\begin{document}

\date{accepted to MNRAS, November 28 2014}

\pagerange{\pageref{firstpage}--\pageref{lastpage}} \pubyear{2013}

\maketitle

\label{firstpage}
%###########################################################################################################

\begin{abstract}
General relativistic magnetohydrodynamic (GRMHD) simulations of accreting black holes in the radiatively inefficient regime show that systems with sufficient magnetic poloidal flux become magnetically arrested disc (MAD) systems, with a well-defined relationship between the magnetic flux and the mass accretion rate. Recently, \cite{Zamaninasab2014} report that the jet magnetic flux and accretion disc luminosity are tightly correlated over 7 orders of magnitude for a sample of 76 radio-loud active galaxies, concluding that the data are explained by the MAD mode of accretion. Their analysis assumes {\it radiatively efficient} accretion, and their sample consists primarily of radiatively efficient sources, while GRMHD simulations of MAD thus far have been carried out in the {\it radiatively inefficient} regime. We propose a model to interpret MAD systems in the context of multiple accretion regimes, and apply it to the sample in \cite{Zamaninasab2014}, along with additional radiatively inefficient sources from archival data. We show that most of the radiatively inefficient radio-loud galaxies are consistent with being MAD systems. Assuming the MAD relationship found in radiatively inefficient simulations holds at other accretion regimes, a significant fraction of our sample can be candidates for MAD systems. Future GRMHD simulations have yet to verify the validity of this assumption.
\end{abstract}


\section{Introduction}\label{sec:intro}

Magnetically arrested disc (MAD) is a model for accretion flow around black holes (BHs) where the magnetic field becomes dynamically important \citep{1974Ap&SS..28...45B,2003ApJ...592.1042I,2003PASJ...55L..69N}. According to the MAD model, systems with sufficient initial poloidal magnetic flux evolve to a state in which the magnetic field threading the BH saturates to a maximal level, and disrupts the accretion flow. Recent general relativistic magnetohydrodynamic (GRMHD) simulations have confirmed the MAD system \citep{Sasha2011,2012MNRAS.423.3083M}. Such simulations have been carried out in the radiatively inefficient regime. They find that MAD systems with sufficient dimensionless BH spin ($a_*\gtrsim0.5$) can efficiently produce relativistic jets by extraction of energy from the rotating BH via the Blandford-Znajek effect \citep{BlandfordZnajek1977}.

More broadly, GRMHD simulations have produced a wide range of accretion discs with jets \citep{Meier2001,Ohsuga2009,Yuan2014}. In general, the properties of the discs and jets depend on the Eddington accretion rate, black hole spin, and magnetic field configuration. In the low-Eddington accretion limit (where accretion luminosity is typically less than $2$~per cent of the Eddington luminosity \citep{Maccarone2003b}), the discs are optically thin and geometrically thick, termed advection-dominated accretion flows (ADAFs; also referred to as radiatively inefficient accretion flow (RIAF) regime in the literature \citep{Narayan1995,Narayan2012,Yuan2014}). An ADAF system can evolve under standard and normal evolution (SANE) if the disc is initially not threaded with significant poloidal magnetic flux, otherwise it may become a MAD system if enough poloidal flux exists. At higher Eddington accretion rates, the accretion disc becomes a geometrically thin, radiatively efficient disc \citep{1973A&A....24..337S}. The long-term thermal stability of such discs \citep{2009ApJ...691...16H, 2013ApJ...778...65J} are debated, meaning that systems found in this regime could possibly be in a transitory state to another regime. At accretion rates higher still, near the Eddington limit or super-Eddington, the accretion flow becomes geometrically thick and optically thick, and is described by the slim disc model \citep{1988ApJ...332..646A}. Recent simulations in this regime have revealed steady state solutions with jets \citep{2014MNRAS.439..503S,2014arXiv1407.4421S,2014MNRAS.441.3177M}, and jet kinetic power is found to be affected strongly by the black hole spin. 

Efforts are made to characterize the accretion modes of observed accreting black hole (BH) systems by investigating correlations between the jet and disc properties and comparing to predictions from theory. \cite{Zamaninasab2014}, hereafter Z14, report a tight correlation between the jet magnetic flux and the accretion disc luminosity for a sample of 76 radio-loud AGNs, and claim the tight correlation is best explained if all the jet-producing supermassive black holes (SMBHs) in their sample are accreting as {\it radiatively efficient} MAD systems. However, MAD systems realized by GRMHD simulations thus far all fall into the {\it radiatively inefficient} regime. This accretion regime actually predicts a different slope from the linear slope observed in Z14. But in fact the linear slope in Z14 is partially biased by a common multiplicative factor of the black hole mass in the two correlated variables. This motivates the present authors to develop a model for MAD systems in various accretion regimes (radiatively inefficient ADAF, radiatively efficient thin disc, and Eddington-saturated slim disc) to characterize the accretion modes of the Z14 sample.

In \S~\ref{sec:model} we model observables of MAD systems in ADAF, thin disc, and slim disc accretion regimes. We test our model against observables in the Z14 sample along with additional archival data of active galaxies. The data sets are summarized in \S~\ref{sec:data}. The results of our analysis are shown in \S~\ref{sec:results}. In \S~\ref{sec:conc} we offer our concluding remarks.

\section{MAD in ADAF, thin disc, and slim disc regimes}\label{sec:model}

The MAD model predicts that the poloidal flux threading the BH, $\Phi_{\rm BH}$, scales with the mass accretion rate $\dot{M}$ and the gravitational radius $r_g = GM/c^2$ in the following fashion \citep{Sasha2011}
\begin{equation}\label{eq:MADpred}
\Phi_{\rm BH} \sim 50\left(\dot{M}r_g^2 c\right)^{1/2}.
\end{equation}
Both sides of this relation may be related to observable quantities, with some reasonable assumptions.

Assuming flux freezing, the BH poloidal flux may be equated with the polodial magnetic flux of the jet $\Phi_{\rm jet}$ (Equation (1) in Z14) as
\begin{equation}\label{eq:LHS}
\Phi_{\rm jet} = 1.2\times 10^{34} f(a_*)\Gamma\theta_{\rm j} \left[\frac{M}{10^{9}M_{\odot}}\right]
\left[\frac{B^\prime_{1\rm pc}}{1G}\right],
\end{equation}
where $f(a_*)=\left(1+(1-a_*)^{1/2}\right)/a_*$ is a function of the mass-normalized BH spin $a_*$, $\Gamma$ is the bulk Lorentz factor of the jet, $\theta_{\rm j}$ is the jet opening angle and $B^\prime_{1\rm pc}$ is the jet's co-moving frame magnetic field strength one parsec downstream from the SMBH obtained from core-shift measurements (see \S~\ref{sec:data} and Z14).

Taking $L_{\rm acc} = \epsilon\dot{M}c^2$, where $L_{\rm acc}$ is the accretion luminosity and 
$\epsilon$ is the radiative efficiency, Z14 derive that the right-hand side of Equation~\eqref{eq:MADpred} can be rewritten as
\begin{align}\label{eq:RHS}
50 \left(\dot{M}r_g^2 c\right)^{1/2}
&= 2.4\times 10^{34} \left[\frac{\epsilon}{0.4}\right]^{-1/2} \nonumber \\ 
& \times \left[\frac{M}{10^9M_{\odot}}\right]
\left[\frac{L_{\rm acc}}{1.26 \times 10^{47}{\rm erg}\ {\rm s}^{-1}}\right]^{1/2}\ {\rm G}\ {\rm cm}^2. 
\end{align}

Z14 assume that the radiative efficiency $\epsilon$ equals to the accretion efficiency $\eta$. The accretion efficiency describes how much potential energy, per unit rest mass energy, can be maximally extracted from matter accreting onto a BH. In the standard \cite{Novikov1973} model, $\eta=0.082$ for spin $a_*=0.5$ and increases to $\eta = 0.42$ for maximally spinning BHs. Models predict a spin $a_*\gtrsim 0.5$ is necessary for creating highly relativistic BH jets powered by BH spin \citep{2012MNRAS.423L..55T,2012JPhCS.372a2040T}. Therefore, $\eta$ is assumed to not vary significantly across jet-producing MAD systems. Z14 assume $\epsilon = \eta =  0.4$ for all their sources, meaning all sources are radiatively efficient accreting SMBHs with near-maximal BH spin. 

However, the radiative efficiency $\epsilon$ need not equal the accretion efficiency $\eta$. The radiative efficiency depends on both the accretion efficiency and the nature of the accretion flow. The radiative efficiency and accretion efficiency only coincide for radiatively efficient sources and diverges in the radiatively inefficient regime \citep{Merloni2008}. In particular many powerful radio galaxies fall into this radiatively inefficient regime, with radiative efficiency $\epsilon<10^{-2}$ \citep{Merloni2007}. It is worth pointing out that all GRMHD simulations of MAD systems have been carried out in this regime \citep{Sasha2011,2012MNRAS.423.3083M}, where the Z14 assumption of high-radiative efficiency does not apply. 

Z14 observe a tight linear correlation between $\Phi_{\rm jet}$ and $L_{\rm acc}^{1/2}M$ for all their sources (demonstrated in Figure 2 of their paper), which is a prediction from radiatively efficient MAD models. For radiatively inefficient sources, however, the relation is not expected to be linear, as we will describe in a later part of this section (see discussion around Equations~\eqref{eq:eff} and \eqref{eq:main}). The reason Z14 observe a tight linear correlation is in part due to a mathematical bias: both axes are multiplied by a common factor of BH mass $M$ (Equation \eqref{eq:LHS} and \eqref{eq:RHS}). The mass spans approximately $3$ orders of magnitude, compared to $\Phi_{\rm jet}/M$ which spans only $2$ orders of magnitude, and $L_{\rm acc}^{1/2}$, which spans $4$ orders of magnitude, artificially stretching out the correlation to a linear one. 

\begin{figure}
\centering
\includegraphics[width=0.5\textwidth]{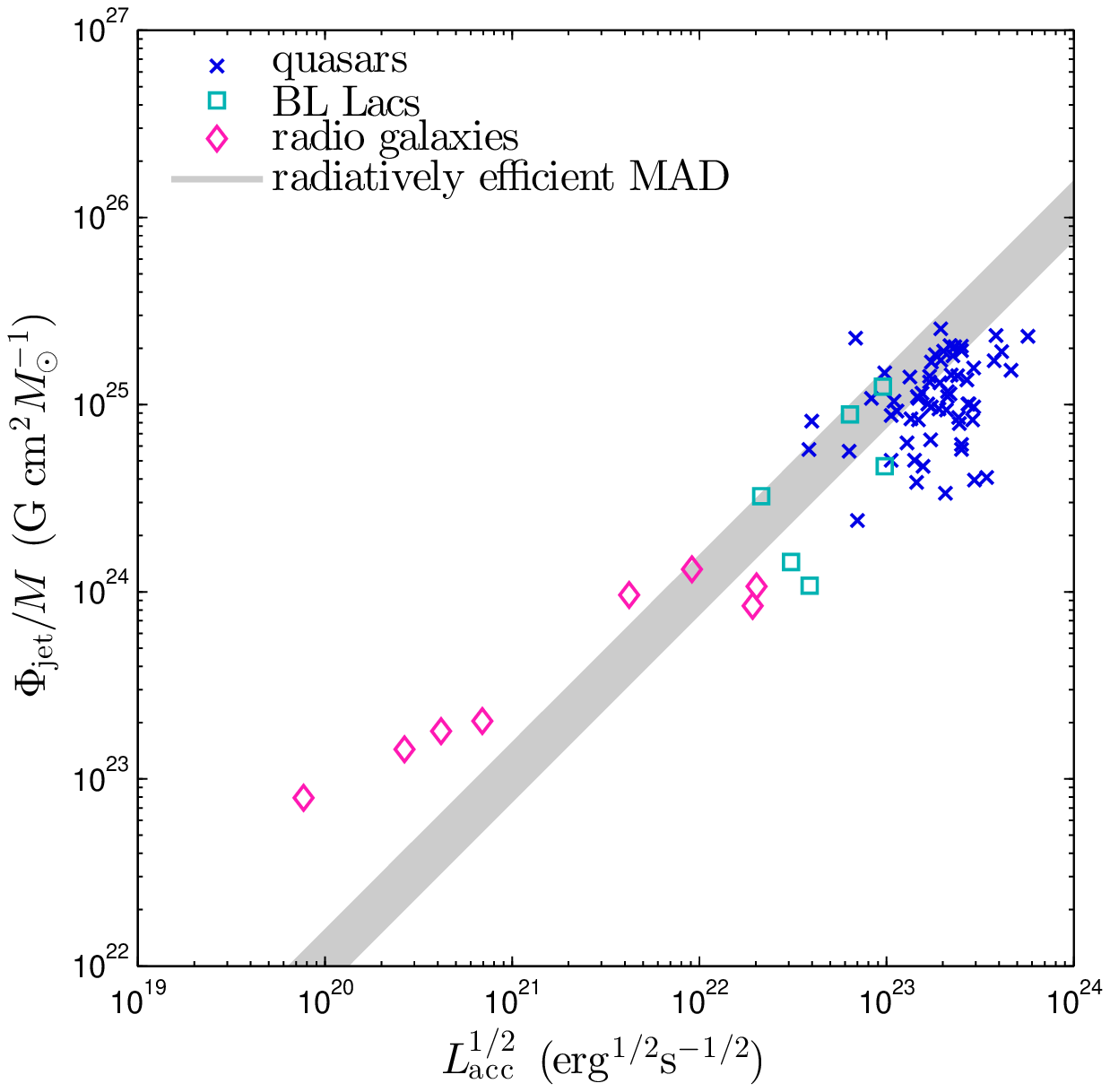}
\caption{The jet magnetic field versus accretion luminosity for the sample of active galaxies in Z14. We repeat the analysis of Z14, but dividing both axes by mass to eliminate the common mass dependence. Note the $y$ axis is just proportional to $B^\prime_{1\rm pc}$, that is, the jet's co-moving frame magnetic field strength one parsec downstream from the SMBH. The extent of the possible parameter space occupied by the radiatively efficient MAD relation for $\eta=0.1$--$0.4$ is shown in the shaded grey region. The data is not consistent with all sources being radiatively efficient MAD. The radio galaxies in the low-luminosity end have too strong magnetic fields. Some quasars have too weak magnetic fields.}
\label{fig:BvsL}
\end{figure}

A simple re-analysis on the data may be performed by dividing through both sides of the relation by mass. We show where the objects (quasars-- blue crosses, BL Lacs  [referred to as blazars in Z14]-- turquoise squares, and radio galaxies-- magenta diamonds) in the Z14 sample lie in this new parameter space in Fig.~\ref{fig:BvsL}. The prediction of radiatively efficient MAD models is shown with the grey shaded region, which allows $\epsilon$ to vary between $0.1$ to $0.4$ in Equation \eqref{eq:RHS}. Low-luminosity radio galaxies are found to lie above the radiatively efficient MAD relation, while some of the quasars lie below it. 

We now describe MAD systems in the context of ADAF, thin disc, and slim disc accretion regimes, which have different radiative efficiencies. We present here physically-motivated model for radiative efficiencies as a function of accretion rate. The model is an extension of the one presented in \cite{Merloni2008,Mocz2013}; which describes powerful jetted sources falling into a radiatively inefficient and a radiatively efficient regime. Our model is described by three regimes: ADAF, thin disc, and slim disc. In the ADAF regime, the system is an optically thin, geometrically thick accretion flow, which occurs when the accretion rate is well below the Eddington limit. This regime corresponds to the low-kinetic (LK) regime described in \cite{Merloni2008}. The thin disc regime is radiatively efficient, and occurs at larger accretion rates (relative to the Eddington rate), referred to as the high-kinetic (HK) regime in \cite{Merloni2008}. We extend the model to a third regime for Eddington rates that are are close to or above the Eddington limit, to account for the insights from recent slim disc simulations \citep{2014MNRAS.439..503S,2014arXiv1407.4421S,2014MNRAS.441.3177M}.\footnote{These simulations have produced systems that appear to extract BH rotational energy through a process similar to the Blandford-Znajek mechanism, implying that MAD is possible in the near- to super-Eddington regime.} In this regime, the system has radiative power saturated near the Eddington luminosity, but it is possible for the kinetic output from jets to be far greater still. The described model also tends to be in agreement with observations of the various X-ray states of X-ray binaries \citep{Maccarone2003,Best2012}, which can serve as low-mass analogues to SMBHs.

In our model, we define $\dot{m}=\eta\dot{M}c^2/L_{\rm Edd}$, the unit-less accretion rate (i.e., normalized by the Eddington luminosity). We relate the Eddington-normalized accretion luminosity $\lambda\equiv L_{\rm acc}/L_{\rm Edd}$ and Eddington-normalized jet power $L_{\rm jet}/L_{\rm Edd}$ to the accretion rate $\dot{m}$. In all three regimes, the jets are assumed to be efficiently extracting energy from accretion: $L_{\rm jet}/L_{\rm Edd}\propto \dot{m}$, where $L_{\rm jet}$ is the jet total kinetic and magnetic energy. We will derive the proportionality constant for MAD systems (Equation~\eqref{eq:main}).

The Eddington-normalized accretion luminosity $L_{\rm acc}/L_{\rm Edd}$, on the other hand, is given by:
\begin{equation}\label{eq:eff}
L_{\rm acc}/L_{\rm Edd} =
\begin{cases}
\lambda_{\rm crit} \left(\frac{\mdot}{\mdot_{\rm crit}}\right)^2 & \mdot\leq \mdot_{\rm crit} \\
\mdot & \mdot_{\rm crit} < \mdot \leq \mdot_{\rm crit,2} \\
\lambda_{\rm slim} & \mdot > \mdot_{\rm crit,2}
\end{cases}
\end{equation}
The regime $\mdot\leq \mdot_{\rm crit}$, where $\mdot_{\rm crit}=\lambda_{\rm crit} = 2\cdot10^{-2}$ \citep{Maccarone2003b} is the radiatively inefficient ADAF regime. Here the disc accretion luminosity scales as $L_{\rm acc}/L_{\rm Edd}\propto \dot{m}^2$. The regime $\mdot_{\rm crit} < \mdot \leq \mdot_{\rm crit,2}$, where we take $\mdot_{\rm crit,2}=1$, is the radiatively efficient thin disc. Here, $L_{\rm acc}/L_{\rm Edd}\propto \dot{m}$. The third regime, $\mdot > \mdot_{\rm crit,2}$, represents slim discs, which are approximately radiating at the Eddington rate, but the total kinetic output can far exceed the Eddington rate. Here we set $L_{\rm acc}/L_{\rm Edd} = \lambda_{\rm slim} = 1$.

The total power of the jets (in the form of relativistic electrons and magnetic field) can be estimated from the magnetic field measurements as \citep{1979ApJ...232...34B}
\begin{equation}\label{eq:BZ}
L_{\rm jet} = 4.5\cdot10^{46} \beta_{\rm j} \left(\frac{B^\prime_{\rm 1pc}}{1~{\rm G}}\right)^2(\Gamma\theta_{\rm j})^2~{\rm erg}~{\rm s}^{-1}
\end{equation}
where $\beta_{\rm j}c$ is the jet velocity.

Our goal is to study the Eddington ratio $\lambda$ as a function of $\mdot$ from our observables, to check the consistency of MAD in this multi-regime accretion model. We obtain a proxy for $\mdot$ for MAD systems. Assuming the MAD model, i.e., equality between Equations~\eqref{eq:LHS} and \eqref{eq:RHS}, and using Equation~\eqref{eq:BZ} for the jet power, we derive the relation
\begin{equation}\label{eq:main}
\dot{m}\eta(a_*)^{-1}f(a_*)^{-2} = 1.75 \frac{L_{\rm jet}}{L_{\rm Edd}} \beta_{\rm j}^{-1}
\end{equation}
The term $\eta(a_*)^{-1}f(a_*)^{-2}$ on the left-hand side is a function of BH spin $a_*$, which is restricted to the range $[0.9,2.9]$  for $a_*\in[0.5,1]$ assuming \cite{Novikov1973} accretion efficiency dependence on BH spin. $\eta(a_*)^{-1}f(a_*)^{-2}=1$ for a spin of $a_*=0.54$.

We remark that we have characterized the accretion state of a SMBH by the Eddington-normalized accretion luminosity, $\lambda$. The transition of the BH accretion state at $\lambda_{\rm crit}$ is originally motivated by observations of BH and neutron star X-ray binaries, which show temporal evolution in their spectral states (i.e., transitions between a thermal-dominated spectrum (`soft') X-ray state and non-thermal powerlaw spectrum (`hard') X-ray state). These spectral states are correlated with the accretion luminosity, making $\lambda$ a good proxy for characterization of the accretion state in X-ray binaries and SMBHs \citep{Maccarone2003,Merloni2008}, though the correlation is not perfect due to hysteresis effects in spectral state transitions \citep{MaccaroneCoppi2003}.

In \S~\ref{sec:results}, we show the analysis of $L_{\rm acc}/L_{\rm Edd}$ versus $\dot{m}\eta(a_*)^{-1}f(a_*)^{-2}$, assuming all sources are MAD, to demonstrate consistency with the picture of multiple accretion regimes (Equation~\ref{eq:eff}).

\section{DATA}\label{sec:data}

Z14 collect a sample of $76$ active galaxies with estimates of jet magnetic field strength, accretion luminosity, and BH mass from the literature (see Z14 and references therein). The sample consists of quasars (mostly flat spectrum radio quasars), BL Lacs, and radio galaxies. BH mass is obtained from the widths and luminosities of optical broad lines and accretion luminosities from the broad line luminosities \citep{Liu2006,Torrealba2012,Shaw2012,Palma2011,Woo2002}. Measurements of the magnetic fields of the jets come from recent core-shift measurements \citep{Pushkarev2012}. The frequency-dependent shift in the location of the radio core can be used to estimate the magnetic field strength and electron number density, assuming a conically-shaped flow, small jet opening angle, and constant bulk velocity \citep{Lobanov1998,Hirotani2005}. It has been assumed that these sources are viewed at their critical angle so that the Doppler factors can be estimated by the apparent velocity of the jet, and this assumption has been tested on a subset of sources with known Doppler factors and viewing angles (\cite{Pushkarev2012}, Z14). A small number of radio galaxies are also included in the sample, with individual core-shift measurements from several papers \citep{Lobanov1998,2004A&A...426..481K, 2008A&A...483..759K, 2011A&A...530L..11M, 2011Natur.477..185H,2011A&A...533A.111M}.

We supplement the sample with $48$ active galaxies (mostly radiatively inefficient) from other data sets which derive jet power using techniques other than core-shift measurements. We include data from \cite{Merloni2007} and \cite{Russell2013} who analysed the jet kinetic power for radiatively inefficient active galaxies. These kinetic powers were obtained from estimated $p\,dV$ work done to inflate the cavities and bubbles observed in the hot X-ray emitting atmospheres of the host galaxies and clusters. These estimates of jet power are averages over the buoyant rise time of a bubble ($10^7$--$10^8$ years). \cite{Russell2013} also measured a few sources with weak jets. These studies provide estimates of the kinetic jet power, accretion luminosity, and BH mass. In addition, we include the radio galaxies from the study by \cite{Godfrey2013}, who use observed parameters of the jet terminal hotspot (hotspot size and equipartition magnetic field strength) to derive jet power. Accretion luminosity for most of the sources in \cite{Godfrey2013} have been estimated by \cite{Hardcastle2009} by modelling the X-ray and mid-IR spectra of these sources. Three sample objects (3C 98, 3C 33, 3C 223) in \cite{Godfrey2013} have BH mass estimates from observed stellar velocity dispersion measurements \citep{Woo2002}.

We note that magnetic field measurements using the core-shift effect assume a single-component, conical jet \citep{Pushkarev2012}. Multi-component models of horizontal jet structure have been proposed to explain spectral energy distributions of BL Lac objects and radio galaxies, such as the spine-sheath model \citep{Swain1998,Ghisellini2005}. This model assumes that the jet consists of two components: a highly relativistic inner spine, and a slower (still relativistic) outer layer. Observed relationships in the spectra versus the luminosity for blazar jets as a function of beaming angle suggest that they may have some horizontal structure \citep{Meyer2011}. Thus our estimates of the jet power from the magnetic fields measured by the core-shift effect may have some degree of bias due to the assumption of a single-component jet. However, the degeneracies due to the many free parameters of the spine-sheath model are hard to constrain, even given detailed observations of the spectral energy distribution \citep{Ghisellini2005}, thus we do not attempt to apply the model to the radio core-shift measurements. We remark that we have included sources with estimates of jet power that are independent of horizontal jet structure, such as estimates from $p\,dV$ work in X-ray cavities. As we will show in \S~\ref{sec:results}, these sources and the sources with jet powers estimated using a single-component jet assumption exhibit the same correlation between jet power and accretion luminosity (Fig.~\ref{fig:lambdavsmdot}).

\section{Results}\label{sec:results}

\begin{figure}
\centering
\includegraphics[width=0.5\textwidth]{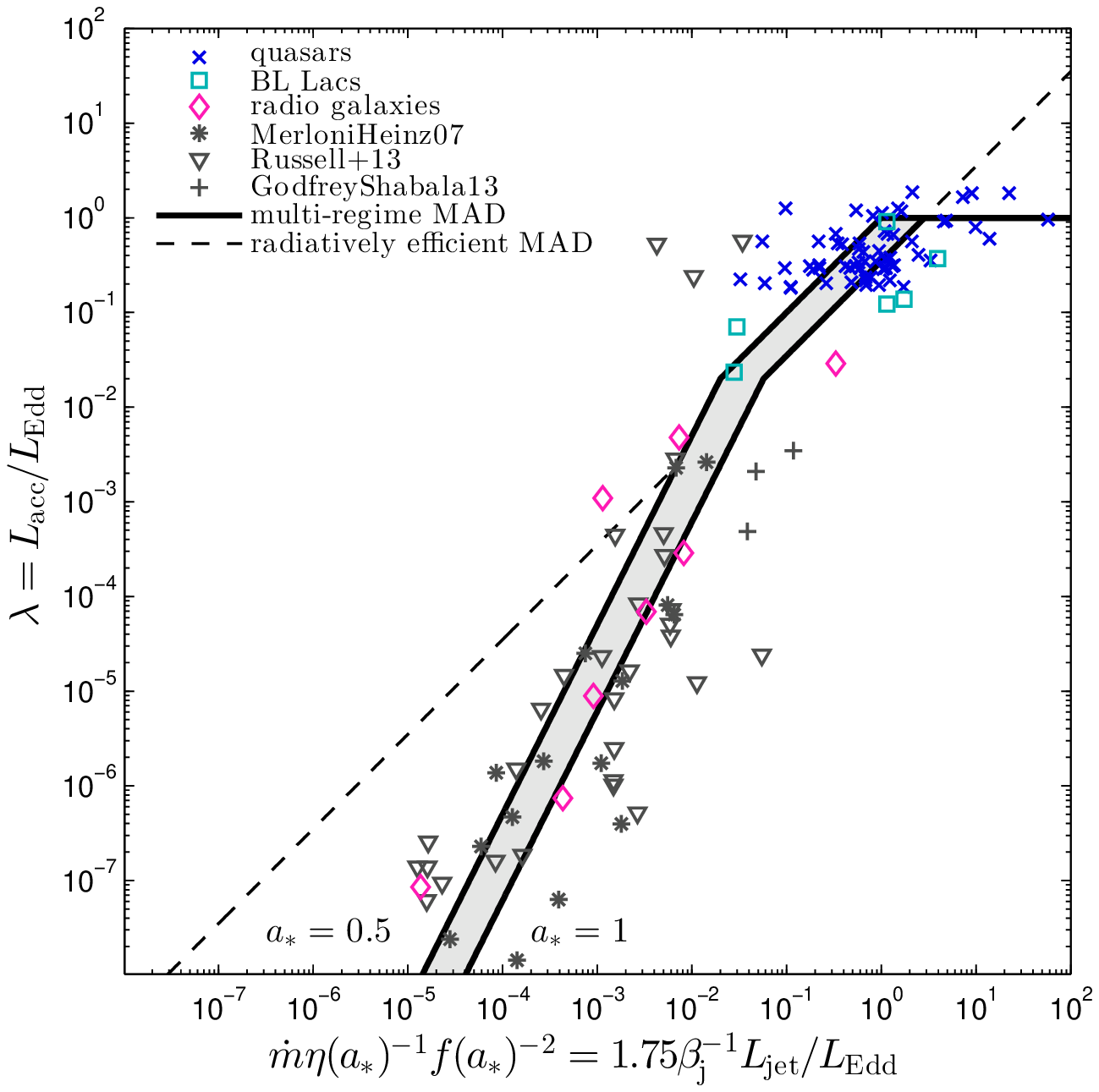}
\caption{Eddington normalized radiative versus kinetic output for our sample of active galaxies. We supplement the data set from Z14 (represented by the same symbols as in Fig.~\protect\ref{fig:BvsL}) with $48$ active galaxies (mostly radiatively inefficient sources) from the literature (grey symbols). The shaded grey area represents the predicted region MAD systems with spins $a_*\in[0.5,1]$ in the ADAF, thin discs, and slim disc regimes are expected to lie. Most of the sources are consistent with the model. Some of the scatter may be accounted for by intrinsic variability in the MAD relation (in Equation~\protect\eqref{eq:MADpred}, the proportionality constant of $50$ can fluctuate by factor of $\sim 2$). The radiatively efficient MAD relation with maximum BH spin, assumed in Z14, is plotted for comparison (dashed-line).}
\label{fig:lambdavsmdot}
\end{figure}

In Fig.~\ref{fig:lambdavsmdot} we present the parameter space occupied by MAD systems in ADAF, thin disc, and slim disc regimes, and show where the active galaxies in our sample lie in the parameter space. On the $x$ axis, we have $\dot{m}\eta(a_*)^{-1}f(a_*)^{-2}$, and on the $y$ axis we have $L_{\rm acc}/L_{\rm Edd}$. We obtain $\dot{m}\eta(a_*)^{-1}f(a_*)^{-2}$ from the observations using  Equation~\eqref{eq:BZ}, assuming $\theta_{\rm j}=1/\Gamma$. Note that $\dot{m}\eta(a_*)^{-1}f(a_*)^{-2}$ then becomes only proportional to $(B^\prime_{\rm 1pc})^2$, as dependence on  $\beta_{\rm j}$ cancels out. The data sets other than Z14 give direct estimates of $L_{\rm jet}$ instead of $B^\prime_{\rm 1pc}$, so in these cases we assume highly-relativistic jets: $\beta_{\rm j}=1$. For different values of BH spin $a_*\in[0.5,1]$, MAD in the context of multi-mode accretion regimes (Equation~\eqref{eq:eff}) predicts slightly different broken power-law relations in the parameter space. We plot the MAD predictions with BH spins at the two extreme values in solid black lines, which enclose the possible occupied parameter space for MAD systems, shown as the shaded grey region. For comparison, we plot the radiatively efficient MAD relation with maximum BH spin, as was assumed in Z14 (dashed-line).

We find that the majority of the systems lie within the region predicted by MAD in the three different accretion regimes we consider, instead of all being consistent with a single, radiatively efficient, regime. The data show scatter around the MAD prediction. Some of the scatter may be accounted for by intrinsic variability in the MAD relation: in Equation~\eqref{eq:MADpred}, the proportionality constant of $50$ can fluctuate by factor of $\sim 2$ in simulations \citep{Sasha2011}. In addition, further scatter is expected from uncertainties in the BH mass measurements.

The radio galaxies in Z14, which were classified as outliers in our reanalysis of Z14 under radiatively efficient assumptions (pink diamonds in Figure~\ref{fig:BvsL}), are consistent with MAD predictions in the radiatively inefficient regime. The Z14 sample does not include many sources in this regime, so we have supplemented our sample with a large number of radio galaxies from archival data, as described in \S~\ref{sec:data}. These systems are also found to agree well with MAD predictions in the radiatively inefficient regime. We stress again it is in this regime where MAD systems have been thus far realized by GRMHD simulations. 

The quasars and BL Lacs in Z14, most of which are in the radiatively efficient regime, agree with the MAD prediction in the said regime. We note GRMHD simulations have yet to show that MAD systems are achievable in this regime, and that Equation~\eqref{eq:MADpred} remains valid. $40$ of the $68$ quasars in the Z14 sample have accretion rates that fall within a factor of $2$ of the MAD prediction (Fig.~\ref{fig:lambdavsmdot}). A fraction of the Z14 sample quasars lie to the left of the MAD prediction, meaning that their jets are too weak to be MAD systems. In addition, the radiatively efficient sources from \cite{Russell2013} produce jets $2$ orders of magnitude weaker than the MAD prediction (grey triangles with $\lambda>0.02$ in Fig.~\ref{fig:lambdavsmdot}). Such sources are likely to be radio-quenched active galaxies with weak jets, a behaviour that is observed for a population of accreting SMBHs with $\lambda>\lambda_{\rm crit}$ \citep{Maccarone2003,Russell2013}. Analogous behaviour is also observed in BH X-ray binaries, where BHs with $0.01\lesssim \lambda\lesssim 0.1$ do not exhibit powerful jets \citep{Tananbaum1972, Harmon1997,Fender2004rev} (note that here the fundamental correlation is with spectral state, and that the spectral state correlates with accretion luminosity). These radio-quenched active galaxies are unlikely to be MAD systems. In fact, we expect more of these systems with even weaker jets could populate the region further left from the grey triangles. The absence of them in Figure~\ref{fig:lambdavsmdot} is likely due to the selection effect that quasars with weaker jets are more difficult to detect with radio observations.

We also observe evidence that some sources fall into the slim disc regime: there is the clear trend that $\lambda$ saturates near $1$, while the Eddington-normalized jet power reaches up to $100$, in agreement with slim disc picture where the radiative output is Eddington limited but the jet power output can be much larger. 

We remark that in our analysis, both axes are multiplied by inverse BH mass. However, we do not expect this to significantly bias our result. The inverse BH mass spans only $3$ orders of magnitude, while $L_{\rm acc}$ spans $7$ orders of magnitude and $\beta_{\rm j}^{-1}L_{\rm jet}$ spans $5$ orders of magnitude. More importantly, the observed correlation is a broken power-law, rather than a linear relation (which would have been the case if the result is biased by the common multiplication of inverse mass), implying that with this combination of observable parameters we are recovering an underlying physical trend, not a mathematical artefact. 

\section{Concluding Remarks}\label{sec:conc}
Accretion around BHs occurs in three regimes: ADAF, thin disc, and slim disc. ADAFs are radiatively inefficient, thin discs are radiatively efficient, and slim discs have accretion luminosities saturated at the Eddington limit. GRMHD simulations in the {\it radiatively inefficient} regime have shown that accreting BHs with sufficient poloidal manetic flux become MAD systems. Z14 analysed a sample of $76$ active galaxies, with the assumption that all sources are {\it radiatively efficient}, and claimed all objects are consistent with MAD. Under the radiatively efficient assumption, MAD predicts a linear correlation between $\Phi_{\rm jet}$ and $L_{\rm acc}^{1/2}M$, which was observed in the Z14 sample. However, the linear correlation observed was partially biased by a common dependence on BH mass in both variables. The radiatively inefficient MAD systems are predicted to follow a different slope. We propose a model to relate observables of MAD systems operating in ADAF, thin disc, and slim disc regimes. We show the majority of the Z14 sample is actually consistent with MAD systems spanning three accretion regimes instead of all being radiatively efficient. The Z14 sample does not contain many radiatively inefficient sources, therefore we collected more archival data and show that powerful radio galaxies are consistent with MAD systems in the radiatively inefficient regime, which is the regime where current GRMHD simulations of MAD systems are most directly applicable. Some of the observed systems in the other two regimes (thin disc, and slim disc) are also consistent with MAD, although GRMHD simulations have yet to realize a MAD-like system in these regimes. The characterization of the radiatively inefficient sources in our study nicely complements the recent study by \cite{Ghisellini2014} who showed a large sample of active galaxies with radiatively efficient accretion discs have accretion luminosities and jet powers (measured by the $\gamma$-ray luminosity) consistent with MAD systems.

\section*{Acknowledgements}
The authors would like to thank Ramesh Narayan and Alexander Tchekhovskoy for helpful discussions, and Andy Fabian, Pierre Christian, Fernando Becerra, and the anonymous referee for valuable comments on the manuscript. This material is based upon work supported by the National Science Foundation Graduate Research Fellowship under grant no. DGE-1144152 (PM) and NASA grant NNX14AB47G (XG).

\bibliography{mybib}{}

%###########################################################################################################
\bsp
\label{lastpage}
\end{document}